\documentstyle[epsfig]{camera}
\newcommand{\AmS}{{\protect\the\textfont2
  A\kern-.1667em\lower.5ex\hbox{M}\kern-.125emS}}

\newcommand{\beq}{\begin{equation}}
\newcommand{\eeq}{\end{equation}}
\newcommand{\bea}{\begin{eqnarray}}
\newcommand{\eea}{\end{eqnarray}}

\def\dm2{\Delta m^2}
\def\sq2{sin^2(2\Theta)}

\input epsf

\begin{document}

%%%%%%%%%%%%%%%%%%%%%%%%%%%%%%%%%%%%%%%%%%%%%%%%%%%%%%%%
% The title, all uppercase; if you want to split it in
% two or more lines, put a \\ macro at each line break
% example:
%   \title{TITLE: FIRST LINE\\ SECOND LINE}
%
\title{NEUTRINO ASTRONOMY WITH THE ICECUBE OBSERVATORY AND IMPLICATIONS FOR ASTROPARTICLE PHYSICS}

%%%%%%%%%%%%%%%%%%%%%%%%%%%%%%%%%%%%%%%%%%%%%%%%%%%%%%%%
% The Author(S), Separated By Commas; Do not put a
% comma before the last author, use instead the \And
% macro which produces a normal ``and'' in the
% caps/small caps context
%
\author{PAOLO DESIATI\\on behalf of the ICECUBE COLLABORATION\footnote{http://www.icecube.wisc.edu/collaboration/authorlists/2008/4.html}}

%%%%%%%%%%%%%%%%%%%%%%%%%%%%%%%%%%%%%%%%%%%%%%%%%%%%%%%%
%
\organization{University of Wisconsin - Madison\\Madison, WI, U.S.A.\\{\tt desiati@icecube.wisc.edu}}

\maketitle

\begin{abstract}
The IceCube Observatory is a km$^3$ neutrino telescope currently under construction at the geographic South Pole. It will comprise 4800 optical sensors deployed on 80 vertical strings between 1450 and 2450 meters under the ice surface. Currently IceCube is operational and recording data with 40 strings (i.e. 2400 optical sensors). The IceCube Observatory will collect an unprecedented number of high energy neutrinos that will allow us to pursue studies of the atmospheric neutrino flux, and to search for extraterrestrial sources of neutrinos, whether point-like or unresolved. IceCube results will have an important impact on neutrino astrophysics, especially if combined with observations done with other cosmic messengers, such as $\gamma$ rays or ultra high energy cosmic rays. They may also reveal clues on the origin of cosmic rays at ultra high energies. Here we report results from AMANDA and the most recent results from the first 22 strings of IceCube.
\end{abstract}
\vspace{1.0cm}

\section{Introduction}
\label{s:intro}

The IceCube Observatory is the first km$^3$-scale neutrino telescope under construction (Abbasi et al., 2008b and Klein et al., 2008a) and it is coming into operation in an exciting period for astroparticle physics along with other future large scale projects such as KM3NeT (Katz, 2006).

The quest for understanding the mechanisms that shape the high energy Universe is currently taking many paths.  Since recently $\gamma$ ray astronomy is going through a series of prolific experimental observations with the modern imaging air Cherenkov telescopes, such as H.E.S.S. (van Eldik, 2008 and Volpe, 2008), VERITAS (Hanna, 2008) and MAGIC (Bartko, 2007) and with water Cherenkov detectors such as the Milagro Gamma Ray Observatory (Abdo et al., 2007). The recent detection of TeV $\gamma$ rays from point-like and extended sources, along with their correlation to observations at other wavelengths, is providing grounds for a significant improvement of our knowledge about the nature of galactic and extra-galactic active objects. The increasing number of newly discovered galactic sources and the measurement of their $\gamma$ ray emission energy spectrum is raising important questions about the mechanisms that produce those high energy $\gamma$ rays. Supernov{\ae} are believed to be the sources of galactic cosmic rays, nevertheless the $\gamma$ ray observations from Supernova Remnants (SNR) still do not provide us with a definite and direct evidence of proton acceleration. The competing inverse Compton scattering of directly accelerated electrons may significantly contribute to the observed $\gamma$ ray fluxes, provided that the magnetic field in the acceleration region does not exceed 10 $\mu$G (Gabici and Aharonian, 2007).

Ultra High Energy Cosmic Rays (UHECR) astronomy, initiated by AGASA (Teshima, 2001) and HiRes (Belz, 2007), and currently pursued by the Pierre Auger Observatory (Abraham, 2007), will potentially yield the next breakthrough in astroparticle physics. The identification of sources of cosmic rays will provide a unique opportunity to probe the hadronic acceleration models currently hypothesized. On the other hand, cosmic ray astronomy is only possible at energies in excess of 10$^{19}$ eV, where the cosmic rays are believed to be extragalactic and point back to their sources for which plausible candidates seem to be Gamma Ray Bursts (GRB) and Active Galactic Nuclei (AGN). TeV $\gamma$ rays from those sources are likely absorbed during their propagation between the source and the observer : at $\sim$10 TeV $\gamma$ rays have a propagation length of about 100 Mpc, while at $\sim$100 GeV $\gamma$ rays can propagate much deeper through the Universe.

If the sources of UHECR are the same as the sources of TeV $\gamma$ rays, then proton acceleration is the underlying mechanism and high energy neutrinos are produced by charged pion decays. Neutrinos would provide an unambiguous evidence for hadronic acceleration in both galactic and extragalactic sources, and are the ideal cosmic messengers, since they can propagate through the Universe undeflected and with practically no absorption. But the same reason that makes neutrinos ideal messengers makes them also difficult to detect.

In section \ref{s:gal} we briefly describe the connection between neutrino astronomy and galactic cosmic rays. In section \ref{s:exgal} we address the topic of extragalactic cosmic rays and their connection to cosmogenic neutrinos. Section \ref{s:icecube} reports the recent results obtained by the IceCube Observatory.

\section{Galactic Cosmic Rays}Ê
\label{s:gal}

\begin{figure}[h!]  %%% FIGURE 1 %%%
\centerline{\psfig{file=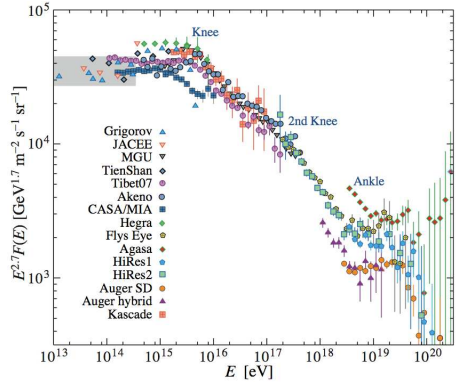,width=6.cm}}
%\epsfysize=5cm %\hspace{4.0cm}
%\epsfbox{./Figure/CR_pd.eps}
\vspace{0.1cm}
\caption[h]{Spectrum of the Cosmic Rays (courtesy of PDG, from T.Gaisser and T.Stanev).}
\label{fig:cr}
\end{figure}

The observed cosmic rays spectrum (see fig. \ref{fig:cr}) has a power law shape up to the so-called knee, at an energy of $\sim$10$^{15}$ eV (i.e. $\sim$ PeV). Up to at least this energy cosmic rays are thought to be generated by Supernova explosions and accelerated via Fermi mechanism by the shock fronts that form in the SNRs within our Galaxy. Shock front acceleration has already been observed in our solar system as driven by solar flares (see Benz, 2008 and references therein). Whatever the sources of galactic cosmic rays are, they must be able to accelerate particles to at least PeV energies.

Diffusive shock acceleration mechanisms allow protons to reach such energies if we assume relatively strong amplification of the magnetic field in the upstream region. An unambiguous indication of proton acceleration to PeV energies by SNRs (the so-called PeVatrons, see Gabici and Aharonian, 2007) would be the detection of $\gamma$ rays at energies in excess of about 100 TeV without cutoff. At these energies the inverse Compton scattering is suppressed and the pionic origin of observed $\gamma$ rays would the most likely interpretation. The complication is that protons can be accelerated to the highest possible energies only when the SNR shock velocity is high enough to allow sufficiently high acceleration rate. This happens only during the first thousand years after the Supernova explosion, while at later times the shock velocity decreases and so does the maximum energy of the accelerated protons. This could explain why the $\gamma$ ray spectrum of SNR RX J1713.7-3946 measured by H.E.S.S. (Aharonian et al., 2007), which extends almost to $\sim$100 TeV, shows a cutoff and becomes rather steep above 10 TeV.

\begin{figure}[h!]  %%% FIGURE 1 %%%
\centerline{\psfig{file=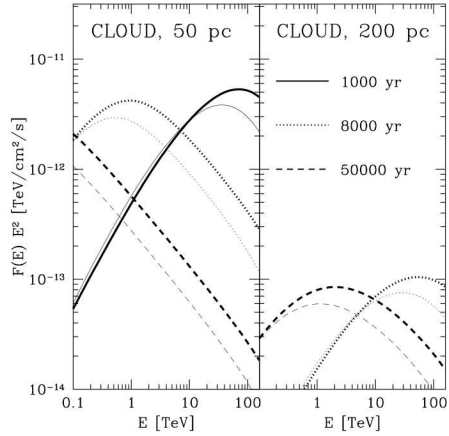,width=6.cm}}
%\epsfysize=5cm %\hspace{4.0cm}
%\epsfbox{./Figure/CR_pd.eps}
\vspace{0.1cm}
\caption[h]{Left: Gamma ray (thick lines) and neutrino (thin lines) spectra from a cloud located at 
50 (left) and 200 pc (right) from the SNR. Different curves refer to different times after the 
explosion (from Gabici and Aharonian, 2007).}
\label{fig:spectra}
\end{figure}

The most energetic particles are produced sooner and can escape the SNR shell earlier than those at lower energies. Since most SNRs are expected to be surrounded by molecular clouds, especially in star forming regions, PeV particles hitting those clouds might produce $\sim$100 TeV delayed $\gamma$ rays. The delay depends on the distance between the SNR and the clouds, whose size would make $\gamma$ ray emission last longer and, therefore, more likely to be observed. This possibility is very appealing and might make it possible to associate multi-TeV $\gamma$ ray observations from molecular clouds to nearby SNRs that acted, in the past, as effective PeVatrons. Under the plausible hypothesis of hadronic origin of TeV $\gamma$ rays from molecular clouds surrounding ancient SNR, Gabici and Aharonian, 2007, estimated the expected neutrino emission (see fig. \ref{fig:spectra}).

A recent survey of the Galactic Center (GC) by H.E.S.S. shows TeV $\gamma$ ray emission corresponding to molecular clouds extending to distances of the order of tens of parsecs around SNR Sgr A East (tagged also HESS J1745-290, see Aharonian et al., 2006). The correlation of $\gamma$ ray emission with target matter at the GC is a strong indication of pionic decay by cosmic rays produced in the nearby SNR (even if the measured $\gamma$ ray energy is only up to about 10 TeV). The measured $\gamma$ ray spectrum implies a spectral index of the cosmic ray spectrum near the GC of about 2.3, which is harder than the one in the solar neighborhood and closer to the intrinsic cosmic ray spectral index. The harder spectrum supports the notion that the observed $\gamma$ rays might be produced by cosmic rays emitted nearby and not yet affected by the spectral softening effect of diffusive propagation.

\begin{figure}[h!]  %%% FIGURE 1 %%%
\centerline{\psfig{file=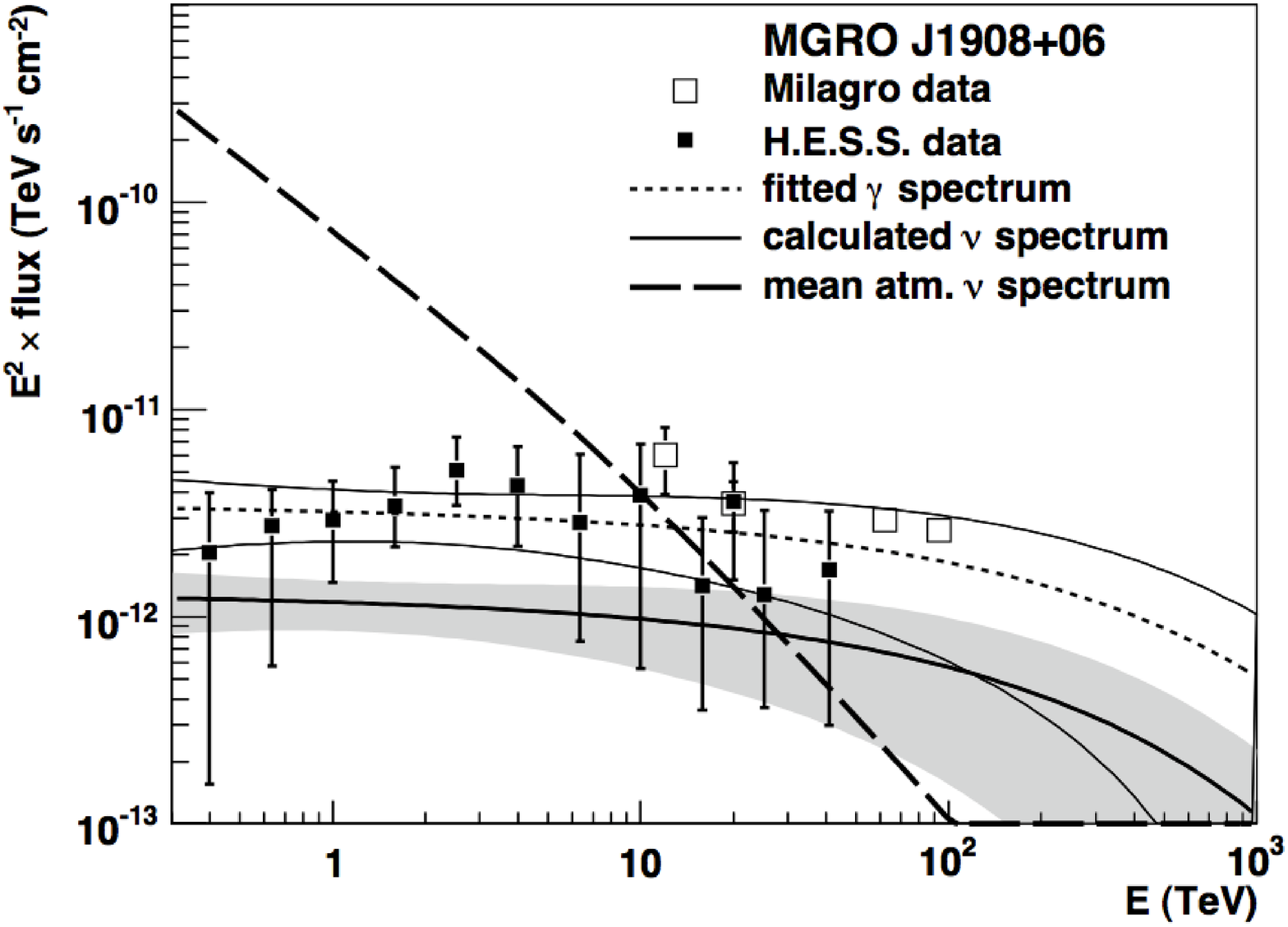,width=7.cm}}
%\epsfysize=5cm %\hspace{4.0cm}
%\epsfbox{./Figure/CR_pd.eps}
\vspace{0.1cm}
\caption[h]{The $\gamma$ ray and neutrino fluxes from MGRO J1908+06. The shaded regions surrounding the fluxes represent the range 
in the spectra due to statistical and systematic uncertainties. Also shown is the flux of atmospheric neutrinos at the same zenith angle as the source (dashed line), taking into account the source size and IceCube angular resolution (from Halzen, Kappes and \'O Murchadha, 2008).}
\label{fig:milagro}
\end{figure}

Milagro, with its global sky survey, observed six extended sources in the Galactic Plane, which are interesting possible candidates of cosmic ray acceleration. In particular, MGRO J1908+06, which had been already observed by H.E.S.S. up to about 40 TeV (Djannati-Atai et al., 2007), was also measured by Milagro up to about 100 TeV with no evidence for an energy cutoff (see fig. \ref{fig:milagro}). This is suggestive of hadronic acceleration and, under this assumption, the corresponding neutrino flux has been calculated by Halzen, Kappes and \'O Murchadha, 2008.

The detection of high energy neutrinos in correlation with TeV $\gamma$ ray observations from galactic sources, would be the awaited unambiguous evidence of hadronic acceleration as the likely production mechanism of galactic cosmic rays. A km$^3$ scale neutrino telescope could achieve such detection sensitivity after several years of data taking (see section \ref{s:icecube}).

\section{Extra-Galactic Cosmic Rays}
\label{s:exgal}

Cosmic rays above $\sim$10$^{18}$ eV (see fig. \ref{fig:cr}) are generally thought to be of extra-galactic origin (Allard et al., 2007). They are hardly confined within the galactic disk by its magnetic field and, also, their energy density is consistent with that emitted by GRB or AGN. Therefore it is believed that those are the sources of the UHECR. Since above 10$^{19}$ eV the cosmic ray deflection in the intergalactic magnetic field is very small, it is possible, in principle, to correlate the direction of the UHECR with known nearby AGNs, especially if protons are the dominant component. The first evidence was published by the Pierre Auger Collaboration, 2007, but the statistical significance is still marginal as is the knowledge of the intensity and structure of the intergalactic magnetic field. Under the assumption that a nearby AGN, such as Centaurus A (Cen A), is a site for hadronic acceleration, that neutrinos and pionic $\gamma$ rays are produced by the underlying p-p interaction, and normalizing to the Pierre Auger Observatory observation, it is estimated that the differential neutrino flux does not exceed $5\times 10^{-13}\left( E\over TeV\right)^{-2}~$TeV$^{-1}$cm$^{-2}$s$^{-1}$ (see Halzen and \'O Murchadha, 2008). Under the assumption that all the AGN that may produce UHECR are like Cen A, the estimation of the diffuse neutrino flux is then about $2\times 10^{-12}\left( E\over TeV\right)^{-2}$TeV$^{-1}$cm$^{-2}$s$^{-1}$sr$^{-1}$. A km$^3$ scale neutrino telescope, therefore, is estimated to be able to detect about 19 events km$^{-2}$yr$^{-1}$ for an E$^{-2}$ spectral shape and about 1 event km$^{-2}$yr$^{-1}$ for E$^{-2.4}$.

\begin{figure}[h!]  %%% FIGURE 1 %%%
\centerline{\psfig{file=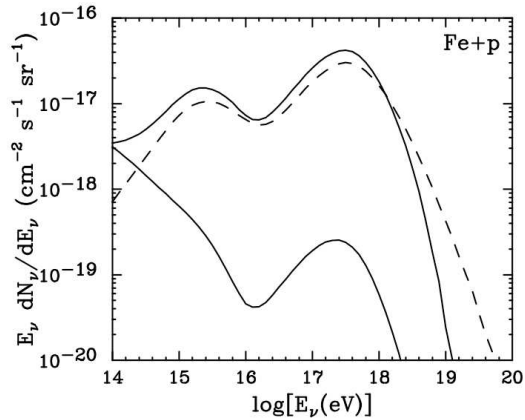,width=7.cm}}
%\epsfysize=5cm %\hspace{4.0cm}
%\epsfbox{./Figure/CR_pd.eps}
\vspace{0.1cm}
\caption[h]{Prediction for cosmogenic neutrino spectra for an assumed mixture of p and Fe UHECR that fit to the Pierre Auger Observatory spectrum and composition measurements. The solid lines denote the models with the highest and lowest rates predicted in a km$^3$ scale neutrino telescope: the lowest line corresponding to 50\% mixture of p and Fe and the highest to only 1\% Fe. The dashed curve is the prediction for an all proton spectrum (from Anchordoqui et al., 2007)}
\label{fig:cosmogenic}
\end{figure}

Experimental observations clearly prove the existence of a UHECR flux suppression above $\sim 6\times 10^{19}$ eV (see fig. \ref{fig:cr}), which is consistent with the GZK cutoff. Above those energies protons interact efficiently with the cosmic microwave and infrared background photons, producing pions which decay to generate a spectrum of ultrahigh energy neutrinos, known as the cosmogenic neutrino flux (see Anchordoqui et al., 2007 and references therein). Considered often to be a guaranteed flux, cosmogenic neutrinos are, in fact, strongly suppressed if a substantial fraction of the UHECR spectrum consists of heavy or intermediate mass nuclei rather than protons. The mass composition of the UHECR is, therefore, an extremely important ingredient in the determination of the expected flux of the cosmogenic neutrinos. If we assume that the UHECR are a mixture of proton and Fe nuclei, the intensity of neutrino-induced muons in a neutrino telescope is estimated to be between about 0.04 km$^{-2}$yr$^{-1}$ and 0.74 km$^{-2}$yr$^{-1}$, for a larger or smaller fraction of Fe in the UHECR, respectively (the corresponding spectra are shown in fig. \ref{fig:cosmogenic}). These event rates are consistent with the Pierre Auger Observatory measurement of the spectrum and composition of the UHECR, if the injected spectrum has an index in the range 1.4-2.1 and if the maximum energy is between 10$^{21}$-10$^{22}$ eV$\cdot $Z (Anchordoqui et al., 2007). Such fluxes are detectable by the IceCube neutrino telescope after several years of data taking.

\section{The IceCube Observatory}
\label{s:icecube}

The IceCube Observatory is a km$^3$ scale neutrino telescope currently under construction at the South Pole (Achterberg et al., 2006 and Klein et al., 2008b) with a total of 4800 Digital Optical Modules (DOM), installed on 80 vertical strings deployed between 1450 and 2450 meters depth in the antarctic ice at the Geographic South Pole. It includes also the surface array IceTop, consisting of a total 80 detector stations, each equipped with two ice Cherenkov tanks, which cover a surface of 1 km$^2$ (see Klepser et al., 2008). The IceCube Observatory is currently halfway instrumented and fully operational. The first 9 in-ice strings (IC-9) recorded data in 2006-07 (see  Achterberg et al., 2007a). The IC-22 data, recorded in 2007-08, and IC-40 data, currently collecting, are presently under analysis. The full array will be completed in 2011. With an energy threshold of about 100 GeV, IceCube is optimized to detect and reconstruct neutrinos of all flavors in the TeV - EeV range. IceCube is currently recording data along with its predecessor AMANDA, which has a smaller volume but a more compacted instrumentation, possibly extending the energy threshold down to about 30-50 GeV.

The AMANDA neutrino telescope has recently analyzed all the data recorded in 3.8 years livetime between 2000 and 2006 (Abbasi et al., 2008a). The selected 6596 muon events from the Northern Sky have average energies between 100 GeV and 8 TeV and a reconstructed track median angular resolution between 1.5$^{\circ}$ and 2.5$^{\circ}$, depending on energy and zenith angle. These events, consistent with the expected atmospheric neutrino flux, have been used to search for galactic and extra-galactic steady point sources of high energy neutrinos, which would manifest as an excess in both direction and energy, since the assumed signal spectral shape is E$^{-2}$. With this spectrum, the analysis is sensitive to possible signal events with energy between 1.9 TeV and 2.5 PeV. The search, performed on a predefined list of 26 galactic and extra-galactic possible sources of high energy neutrinos, including many TeV $\gamma$ ray sources, does not show any significant excess. The highest significance is found in the direction of Geminga with a chance probability (p-value) of 0.0086. The probability of obtaining a p-value less or equal to the measured one for at least one of the 26 sources is 20\%. By stacking the five Milagro TeV $\gamma$ ray sources with the highest pre-trial significance or the highest $\gamma$ ray intensity, a possible correlation with an excess of high energy neutrinos has been searched for. This analysis shows a slight neutrino event excess with chance probability of 20\% and provides an upper limit on the $\nu_{\mu}$ flux per source of 9.7$\times$10$^{-12}\left( E\over TeV\right)^{-2}$ TeV$^{-1}$cm$^{-2}$s$^{-1}$ (90\% CL), without accounting for systematic uncertainties. If we compare this upper flux limit to the predicted neutrino fluxes from fig. \ref{fig:spectra} and  \ref{fig:milagro} we see that the AMANDA sensitivity is about one order of magnitude too low to impact the current predictions. Currently IceCube data are being analyzed to search for steady point sources of high energy neutrinos as well. The IC-22 detector is expected to exceed the 7-year AMANDA sensitivity by a factor of 2 (Finley et al., 2007). The full IC-80 neutrino telescope is expected to achieve a track angular resolution of less than one degree and an order of magnitude improvement over the 7-year AMANDA sensitivity, within three years of operation. The km$^3$ scale neutrino telescope is, therefore, expected to see steady point sources of high energy neutrinos, based on the current estimations.

If individual point-like sources are too faint, then the search for high energy neutrinos from unresolved sources across the Northern Sky might have the most promising discovery potential. The search for muon neutrinos from diffuse sources has been performed with data collected by AMANDA in 807 d livetime between 2000 and 2003 (Achterberg et al., 2007b). Under the hypothesis that the signal has a spectral shape of E$^{-2}$, and since no event excess has been observed at high energy, where the flux of atmospheric neutrinos is low, an upper limit of 7.4$\times$10$^{-11}\left( E\over TeV\right)^{-2}$ TeV$^{-1}$cm$^{-2}$s$^{-1}$sr$^{-1}$ (90\% CL) has been determined within the neutrino energy range of 16 TeV to 2.5 PeV. This upper limit is still more than an order of magnitude higher than the estimated diffuse flux from Halzen and \'O Murchadha, 2008. Once again, the IC-22 detector, the data from which are currently under analysis, is expected to exceed the 4-year AMANDA sensitivity by at least a factor of 2, given the improvement on the event energy estimation and the larger acceptance above 100 TeV. One year of IC-80 is expected to achieve a sensitivity level allowing detection of the diffuse fluxes of neutrinos currently predicted from unresolved AGN.

The search for neutrinos with energy in excess of about 10$^5$ GeV has also been performed in AMANDA, using the experimental data collected between 2000 and 2002 (Ackermann et al., 2008). Since above 10$^{7}$ GeV the Earth is essentially opaque to neutrinos, this search is concentrated on the bright events recorded from the Southern Sky and from the horizon belt. In this case the main background is not represented by the atmospheric neutrinos, but by the large bundles of cosmic muons produced by the impact of high energy cosmic rays with the atmosphere. The search for those bright events did not reveal any significant excess and an upper limit of 2.7$\times$10$^{-10}\left( E\over TeV\right)^{-2}$ TeV$^{-1}$cm$^{-2}$s$^{-1}$sr$^{-1}$ (90\% CL) has been determined within the energy range of 2$\times$10$^5$ GeV and 10$^9$ GeV. This upper flux limit is about one order of magnitude higher than the highest predicted flux of cosmogenic neutrinos from Anchordoqui et al., 2007. IC-80 is expected to improve this sensitivity by at least one order of magnitude for energy below about 10$^9$ GeV within two years of operation..

%\begin{table}[t!]
%\begin{center}
%\caption{}
%\bigskip
%\begin{tabular}{llllll}
%\hline
%geometry & livetime & muons & efficiency & purity & notes\\
%\hline
%IC-9 & 137.4 d & 233 (1.7/d) & $\sim$3\% & $\sim$90\% & published\\
%\hline
%IC-22 & 285 d & $\sim$7,800 (27/d) & $\sim$25\% & $\sim$95\% & preliminary\\
%\hline
%IC-40 & 365 d & $\sim$40,000 (110/d) & {} & {} & prediction\\
%\hline
%IC-80 & 365 d & $\sim$80,000 (220/d) & {} & {} & prediction\\
%\hline
%\end{tabular}
%\end{center}
%\end{table}

%%%%%% Table 2 %%%%%%%%%%%%%%%%%%%%%%%%%%%%%%%%%%%%%%%%%%%%%%%%%%%%%%%%%%%%
%\begin{table}[h!]
%\begin{center}
%\caption{Black holes in the Milky Way Galaxy (Blandford \&
%Gehrels, 1999; Filippenko et al., 1999; Casares, 2001) }
%\bigskip
%\begin{tabular}{llllll}
%\hline
%Source Name  & Identification &Companion&f(M)&$M_{Opt}$  & $M_{BH}$ \\
%             &                &         &   &($M_\odot$)&($M_\odot$)\\
%\hline
%Cygnus X-1   & HD226868  & O9.7 Iab & 0.24   & 24-42        & 11-21\\
%\hline
%\end{tabular}
%\end{center}
%\end{table}
%%%%%%%%%%%%%%%%%%%%%%%%%%%%%%%%%%%%%%%%%%%%%%%%%%%%%%%%%%%%%%%%%%%%%%%%%%%

\section{Conclusions}

Current km$^3$ scale neutrino telescopes under construction such as IceCube, or under design such as KM3NeT, have the potential to uncover the origin of galactic and extra-galactic cosmic rays. Neutrino astronomy, in correlation with observations of TeV $\gamma$ rays and of UHECR, might provide the incontrovertible evidence of hadronic acceleration as the common underlying mechanism that produces the high energy phenomenologies observed with different cosmic messengers. The first generation neutrino telescope AMANDA which proved that neutrino astronomy is feasible, is still providing a very competitive sensitivity for high energies. But it is the new km$^3$ scale neutrino telescopes currently under construction and in operation, that will reach, within a few years of data taking, a sensitivity to possibly assess the current models of hadronic accelerations by SNR GRBs or AGNs. Such observatories have the potential to unveil one of the most intriguing mysteries of the Universe : the origin of the cosmic rays. On the other side, there is the possibility that hadronic acceleration mechanisms suffer more serious inefficiencies than currently hypothesized. In this case even larger scale observatories, providing at least an order of magnitude higher sensitivity at the highest energies, are needed. Such observatories, relying on either radio or acoustic detection, are already under design.

\section{Acknowledgements}
{\small
We acknowledge the support from the following agencies: U.S. National Science FoundationÐOffice of Polar Programs,
U.S. National Science FoundationÐPhysics Division, University of Wisconsin Alumni Research Foundation, U.S.
Department of Energy and National Energy Research Scientific Computing Center, Louisiana Optical Network Initiative
(LONI) grid computing resources, Swedish Research Council, Swedish Polar Research Secretariat, Knut and Alice
Wallenberg Foundation (Sweden), German Ministry for Education and Research (BMBF), Deutsche Forschungsgemeinschaft
(DFG), (Germany), Fund for Scientific Research (FNRS-FWO), Flanders Institute to encourage scientific and
technological research in industry (IWT), Belgian Federal Science Policy Office (Belspo), and the Netherlands Organisation
for Scientific Research (NWO); M. Ribordy acknowledges the support of the SNF (Switzerland); A. Kappes and A.
Gro§ acknowledge support by the EU Marie Curie OIF Program; M. Stamatikos is supported by an NPP Fellowship at
NASAÐGSFC administered by ORAU.
}

%\newpage
%\bigskip
%\bigskip
%{\small
%\noindent {\bf DISCUSSION}

%\bigskip
%\noindent {\bf Lawrence W. Jones:} I was surprised to see you cite the AUGER data
%as supporting a mixed and perhaps p+Fe composition at the highest energies. The
%AUGER publications I have read argued p or p+He, or other very light composition.

%\bigskip
%\noindent {\bf Paolo Desiati:} At ICRC 2007 the AUGER Collaboration showed an apparent
%incerase of the cosmic ray composition above $\sim$3$\times$10$^{19}$ eV. This seems in
%contrast with other observations, but so far it appears a significant result.

%\noindent Anchordoqui et al. claim that they cannot fit p only cosmic ray composition simultaneously
%to the AUGER energy spectrum and composition above 10$^{19}$ eV. They claim the spectrum and
%composition may be compatible with medium mass cosmic rays (i.e. CNO) or with a mixture of p and Fe.
%}

\begin{thebibliography}{9}
{\small
\bibitem{19} Abbasi A. et al. (IceCube Collaboration): 2008a, arXiv:0809.1646
\bibitem{25} Abbasi A. et al. (IceCube Collaboration): 2008b, to be submitted to {\it Nucl. Instrum. Meth.} A, aeXiv:0810.4930
\bibitem{1} Abdo A.A., et al. (Milagro Collaboration): 2007,  {\it Astrophys. J.} {\bf 658} L33
\bibitem{2} Abraham J., et al. (Pierre Auger Collaboration): 2007, {\it Science} {\bf 318} 938
\bibitem{26} Achterberg A., et al. (IceCube Collaboration): 2006, {\it Astropart. Phys.} {\bf 26} 155, arXiv:astro-ph/0604450
\bibitem{21} Achterberg A., et al. (IceCube Collaboration): 2007a, {\it Phys. Rev.} {\bf D 76} 027101
\bibitem{15} Achterberg A., et al. (IceCube Collaboration): 2007b, {\it Phys. Rev.} {\bf D 76}  042008, Erratum-ibid. {\bf D 77} 089904
\bibitem{22} Ackermann M., et al. (IceCube Collaboration): 2008, {\it Astrophys. J.} {\it 675} 1014
\bibitem{13} Aharonian F. et al. (HESS Collaboration): 2006, {\it Nature} {\bf 439} 695
\bibitem{3} Aharonian F. et al. (HESS Collaboration): 2007, {\it Astr. and Astroph.} {\bf 464} 235
\bibitem{4} Allard D., et al.: 2007, {\it Astr. and Astroph.} {\bf 473-1} 59, arXiv:astro-ph/0703633
\bibitem{20} Anchordoqui L.A., et al.: 2007, arXiv:0709.0734
\bibitem{5} Bartko H. (MAGIC Collaboration): 2007 {\it Mod. Phys. Lett.} {\bf A22} 2167
\bibitem{6} Belz J. (HiRes Collaboration): 2007, {\it Nucl. Phys. Proc. Suppl.} {\bf 165} 239. Proceedings of the Cosmic Ray International Seminar (CRIS 2006), Catania, Italy
\bibitem{23} Benz A.O. : 2008, {\it Living Rev. Solar Phys.} {\bf 5} 1
\bibitem{16} Djannati-Atai A. et al. (HESS Collaboration): 2007, in Proceedings of the 30$^{th}$ International Cosmic Ray Conference (ICRC 2007), Merida, Mexico, arXiv:0710.2418
\bibitem{8} Eldik C. van (HESS Collaboration): 2008, {\it Nucl. Instrum. Meth.} {\bf A 588} 72, in Proceedings of the Roma International Conference on Astro-Particle Physics (RICAP'07), Roma, Italy
\bibitem{9} Gabici S. and Aharonian F.: 2007, {\it Astrophys. J. Letters} {\bf 665} L131
\bibitem{28} Finley C., Dumm J. and Montaruli T. (IceCube Collaboration): 2007 in Proceedings of 30th International Cosmic Ray Conference (ICRC 2007), Merida, Mexico, arViv:0711.0353
\bibitem{18} Halzen F., Kappes A. and \'O Murchadha A.: 2008, {\it Phys. Rev.} {\bf D 78} 063004, arXiv:0803.0314
\bibitem{14} Halzen F. and \'O Murchadha A.: 2008, arXiv:0802.0887
\bibitem{10} Hanna D. (VERITAS Collaboration): 2008, {\it Nucl. Instrum. Meth.} {\bf A 588} 26
\bibitem{24} Katz U.F. : 2006, arXiv:astro-ph/0606068 and in these Proceedings
\bibitem{29} Klepser S. et al. (IceCube Collaboration) : 2008, in Proceeding of the 21$^{st}$ European Cosmic Ray Symposium (ECRS 2008 ), Ko{\u s}ice, Slovakia.
\bibitem{7} Klein S. (IceCube Collaboration): 2008a, in Proceedings of the Symposium on Radiation Measurement and Applications (SORMA West 2008), Berkeley, CA U.S.A., arXiv:0807.0034
\bibitem{27} Klein S. (IceCube Collaboration): 2008b, in Proceedings of the 23$^{rd}$ International Conference on Neutrino Physics and Astrophysics (Neutrino 2008), Christchurch, New Zealand, arXiv:0810.0573
\bibitem{17} The Pierre Auger Collaboration: 2007, {Science} {\bf 318} 938
\bibitem{11} Teshima M., et al. (AGASA Collaboration): 2001, in Proceedings of the 27$^{th}$ International Cosmic Ray Conference (ICRC 2001), Hamburg, Germany, Published in Hamburg, 333
\bibitem{12} Volpe F. (HESS Collaboration): 2008, {\it Nucl. Instrum. Meth.} {\bf A 588} 76
}
\end{thebibliography}
\end{document}